\documentclass[aps,prl,floatfix,onecolumn,reprint,amsmath,amssymb,superscriptaddress,longbibliography]{revtex4-1}
\usepackage{amsfonts}
\usepackage{mathrsfs}
\usepackage{amsmath}
\usepackage{color}
\usepackage{natbib}
\usepackage{graphicx}
\usepackage{bm}
\usepackage{amssymb}
\usepackage{amsmath,amssymb,amsfonts,mathrsfs}
\usepackage{stmaryrd}
\usepackage{xspace}
\usepackage{epstopdf}
\usepackage{dcolumn}
\usepackage{longtable}
\usepackage{multirow}
\usepackage{bbding}
\usepackage[colorlinks=true, pdfstartview=FitV, linkcolor=blue, citecolor=blue, urlcolor=blue]{hyperref}

\makeatletter

\newcommand{\Rmnum}[1]{\expandafter\@slowromancap\romannumeral #1@}

\def\iddots{\mathinner{\mkern1mu\raise\p@
\vbox{\kern7\p@\hbox{.}}\mkern2mu
\raise4\p@\hbox{.}\mkern2mu\raise7\p@\hbox{.}\mkern1mu}}

\makeatother
\begin{document}

\title{Anomalous Hall Conductivity as an Effective Means of Tracking the Floquet Weyl Nodes in Quasi-One-Dimensional $\beta$-Bi$_4$I$_4$}
	
\author{Qingfeng Huang}
\thanks{These authors contributed equally to this work.}
\affiliation{Hongshen Honors School of Chongqing University, Chongqing 401331, People's Republic of China}

\author{Shengpu Huang}
\thanks{These authors contributed equally to this work.}
\affiliation{Institute for Structure and Function $\&$ Department of Physics $\&$ Chongqing Key Laboratory for Strongly Coupled Physics, Chongqing University, Chongqing 400044, People's Republic of China}

\author{Tingyan Chen}
\affiliation{Institute for Structure and Function $\&$ Department of Physics $\&$ Chongqing Key Laboratory for Strongly Coupled Physics, Chongqing University, Chongqing 400044, People's Republic of China}

\author{Jing Fan}
\affiliation{Center for Computational Science and Engineering, Southern University of Science and Technology, Shenzhen 518055, China}

\author{Dong-Hui Xu}
\affiliation{Institute for Structure and Function $\&$ Department of Physics $\&$ Chongqing Key Laboratory for Strongly Coupled Physics, Chongqing University, Chongqing 400044, People's Republic of China}
\affiliation{Center of Quantum materials and devices, Chongqing University, Chongqing 400044, People's Republic of China}

\author{Xiaozhi Wu}
\affiliation{Hongshen Honors School of Chongqing University, Chongqing 401331, People's Republic of China}
\affiliation{Institute for Structure and Function $\&$ Department of Physics $\&$ Chongqing Key Laboratory for Strongly Coupled Physics, Chongqing University, Chongqing 400044, People's Republic of China}

\author{Da-Shuai Ma}
\email{mads@cqu.edu.cn}
\affiliation{Institute for Structure and Function $\&$ Department of Physics $\&$ Chongqing Key Laboratory for Strongly Coupled Physics, Chongqing University, Chongqing 400044, People's Republic of China}
\affiliation{Center of Quantum materials and devices, Chongqing University, Chongqing 400044, People's Republic of China}

\author{Rui Wang}
\email[]{rcwang@cqu.edu.cn}
\affiliation{Institute for Structure and Function $\&$ Department of Physics $\&$ Chongqing Key Laboratory for Strongly Coupled Physics, Chongqing University, Chongqing 400044, People's Republic of China}
\affiliation{Center of Quantum materials and devices, Chongqing University, Chongqing 400044, People's Republic of China}
	
\begin{abstract}
While Floquet engineering offers a powerful paradigm for manipulating topological phases, particularly Floquet Weyl semimetals, establishing an experimentally feasible strategy for tracking the dynamic evolution of such states remains a significant challenge.
Here, we propose that the anomalous Hall effect (AHE), as a sensitive, all-electrical probe, can be used to track Floquet Weyl nodes. 
Using first-principles calculations and symmetry analysis on the quasi-one-dimensional material $\beta$-Bi$_4$I$_4$, we demonstrate that circularly polarized light breaks time-reversal symmetry, driving the system from a trivial insulator into a Floquet Weyl semimetal phase characterized by a nonzero Berry curvature flux. 
Crucially, by continuously tuning the polarization phase $\varphi$ of the driving field, we show that the trajectory of the induced Weyl nodes is highly controllable, leading to their migration and eventual annihilation at high-symmetry points. 
We reveal that the anomalous Hall conductivity maps directly onto this topological evolution, serving as a definitive fingerprint for the generation and dynamics of Weyl nodes. 
\end{abstract}
\maketitle


\textit{\textcolor{black}{Introduction. ---}}
In recent decades, the exploration of topological phases of matter has been a significant research focus ~\cite{hasan2010colloquium,qi2011topological,burkov2016topological,wang2017topological,armitage2018weyl,tokura2019magnetic}. 
Topological materials are considered to have broad application potential in the realization of next-generation ultra-low-power devices, making them a frontier field in condensed matter physics and materials science ~\cite{yan2012topological,bernevig2022progress,he2018topological,zhang2009topological}.
Nowadays, rapidly developing theories of band representation, symmetry indices, and topological quantum chemistry have enabled a comprehensive understanding and complete classification of the topological phases of matter
~\cite{po2017symmetry,bradlyn2017topological, kruthoff2017topological, watanabe2018structure, song2018quantitative,zhang2019catalogue,vergniory2019complete,tang2019comprehensive,song2020twisted, xu2021three,elcoro2021magnetic, bouhon2021topological, peng2022topological}.
Based on these facts, achieving function-oriented theoretical design of quantum functional materials through effective control via microscopic mechanisms has become a new topic in the condensed matter physics and materials science communities ~\cite{zunger2018inverse,marzari2021electronic,wang2022inverse}.
Targeted modulation of the symmetry of external fields or perturbations applied to topological materials is one of the effective strategies to achieve the precise function-oriented design of topological states of matter. This approach has been instrumental in numerous experimental and theoretical breakthroughs~\cite{wang2017structural,mutch2019evidence,li2019magnetically,niu2020antiferromagnetic,fan2023complete,tang2023evolution,yang2025sliding}. 

On the other hand, Floquet engineering via periodic driving has emerged as a robust paradigm for manipulating quantum states ~\cite{2013PhysRevLett.110.026603,wang2013observation,2016PhysRevLett.117.087402,hubener2017creating,2018PhysRevLett.120.237403,2018PhysRevB.84.235108,2018PhysRevLett.120.156406,2020flo_deng_deng2020photoinduced,mciver2020light,PRL2022BCL,zhou2023pseudospin,zhan2023floquet,2023PhysRevB.107.085151,PhysRevB.79.081406,add1merboldt2024observationfloquetstatesgraphene}.
Exploiting the correspondence between symmetry and topology, this approach is widely employed to engineer quantum systems with exotic topological, superconducting, and magnetic properties. 
For instance, due to the natural breaking of time-reversal symmetry in circularly polarized light (CPL), Floquet chiral topological superconductivity has been predicted in antiferromagnetic systems with vanishing net magnetic moments~\cite{ning2024flexible}. 
More recently, CPL has been demonstrated as a versatile tool to manipulate spin group symmetries, inducing odd-parity spin splitting in the otherwise spin-degenerate bands of two-dimensional antiferromagnets~\cite{huang2025light,li2025floquet,zhu2025floquet,liu2025light}.
In particular, CPL has been proposed to drive topological insulators, triple-component fermions, and Dirac or nodal-line semimetals into Weyl semimetal phases ~\cite{zhang2016theory,yan2016tunable,chen2018floquet,zhu2020floquet,wang2023floquet,zhan2024perspective,PhysRevB.110.L121118}.
While recent efforts have extended these theoretical models to realistic materials, developing experimentally feasible schemes to track the process of these phase transitions (such as the generation and annihilation of Weyl fermions in Floquet Weyl semimetal) remains a significant challenge.

In the present work, by means of first-principles calculations and symmetry analysis, we propose that detecting the anomalous Hall conductivity is an effective way to track the light-driven topological transition in quasi-one-dimensional $\beta$-Bi$_4$I$_4$, offering a significantly more convenient and experimentally straightforward pathway for manipulating the topological state.
The key idea is to employ a light field with a time-dependent vector potential $\mathbf{A}(t)=\mathrm{A}_{0}\left[\mathrm{cos}\left(\omega t\right), \mathrm{sin}\left(\omega t+\varphi\right),0\right]$ to induce specific symmetry breaking, which directly impacts the topological features of $\beta$-Bi$_4$I$_4$. Here, $\mathrm{A}_{0}$ and $\omega$ are the amplitude and frequency of the incident light. The phase difference $\varphi$ dictates the polarization state: circular for $\varphi=0$, linear for $\varphi=\pi/2$, and elliptical for intermediate values ($0 < \varphi < \pi/2$). When $\varphi=0$, the time-reversal-symmetry-breaking circularly polarized light (CPL) drives $\beta$-Bi$_4$I$_4$ from a normal narrow-gap insulator to a Floquet Weyl semimetal with nonzero anomalous Hall conductivity.
By adjusting $\varphi$ from $0$ to $\pi/2$, the light field gradually becomes LPL with polarization along the [1,-1,0] direction. During this process, the Weyl nodes in the Floquet Weyl semimetal phase gradually move closer and eventually annihilate, accompanied by a decrease in the anomalous Hall conductivity.
When $\varphi=\pi/2$, the time-reversal symmetry of the light-irradiated $\beta$-Bi$_4$I$_4$ is restored, leading to the vanishing of the anomalous Hall effect.
While previous Floquet engineering studies have predominantly relied on modulating the light intensity or frequency of the light to control the Weyl nodes and Fermi arcs, our results reveal that, in $\beta$-Bi$_4$I$_4$, a simple adjustment of the polarization of the driving light suffices to reversibly switch the anomalous Hall effect and track the process of topological phase transition.
The experimental feasibility of the proposed scheme is discussed in detail in the Supplementary Material (SM)~\cite{Hwang}.
 
\textit{\textcolor{black}{Methods. ---}}
Our first-principles calculations were performed using the projector augmented wave method ~\cite{blochl1994projector} implemented in the Vienna \textit{ab-initio} Simulation Package (VASP) ~\cite{kresse1996efficient,kresse1999ultrasoft,kresse1996efficient}. 
The experimental crystal structure of $\beta$-Bi$_4$I$_4$ is adopted with the experimental lattice constants kept fixed throughout the calculations, while the atomic positions are fully optimized ~\cite{von1978wismutmonojodid}. The ion-electron interaction was described by the projector augmented wave (PAW) method ~\cite{PhysRevB.59.1758} with the Perdew-Burke-Ernzerhof (PBE) exchange-correlation functional ~\cite{PhysRevLett.77.3865}.
Due to weak van der Waals (vdW)-type interlayer coupling ~\cite{geim2013van} in $\beta$-Bi$_4$I$_4$, vdW corrections (DFT-D3 method with zero damping, corresponding to IVDW = 11 in VASP) are employed to relax the ionic positions until forces on each ion are less than 0.01~$\mathrm{eV}/$\AA.
To obtain accurate electronic band structures, we used the modified Becke-Johnson (mBJ) exchange potential in conjunction with spin-orbit coupling (SOC).
The kinetic energy cutoff was set to $320~\textrm{eV}$, and the Brillouin zone was sampled with a $6 \times 6 \times 4$ $\Gamma$-centered $k$-mesh ~\cite{monkhorst1976special}.
To investigate the light-manipulated Weyl physics, we constructed a tight-binding Hamiltonian by projecting the Bloch states onto the maximally localized Wannier functions (MLWFs) ~\cite{marzari2012maximally} based on the $p$ orbitals of Bi and I using the wannier90 code.
The surface local density of states and Fermi arcs were calculated via the iterative Green's function method implemented in the WANNIERTOOLS package ~\cite{wu2018wanniertools}.
We simulated the periodic light-matter interaction by introducing the Peierls substitution, $\bm{k} \rightarrow \bm{k} + \mathrm{e}\mathbf{A}(t)/\hbar$ to obtain a time-dependent Hamiltonian ~\cite{hofstadter1976energy}.
In the high-frequency limit, we consider the off-resonant region where the photon energy $\hbar\omega$ is significantly larger than the bandwidth. The time-dependent system is described by an effective static Hamiltonian derived via the Floquet theorem ~\cite{kitagawa2011transport,kitagawa2011transport,bukov2015universal}.
This approach allows us to analyze the light-induced modifications in the system's topological properties without considering the coupling between different Floquet sub-bands. 

\begin{figure}[t]
\includegraphics[width=1\columnwidth]{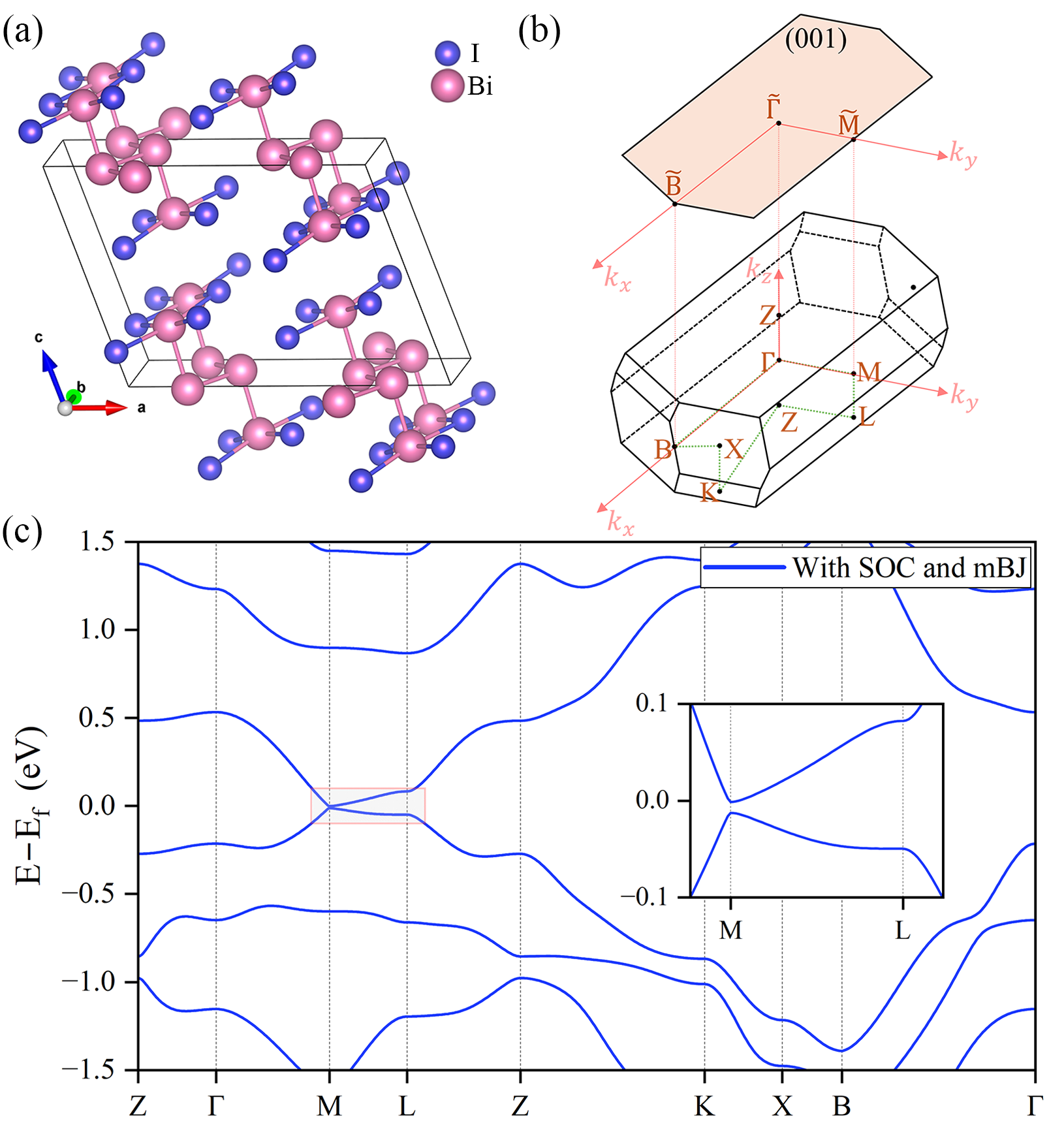}
\caption{\label{fig-1}(a) Crystal structure of $\beta$-Bi$_4$I$_4$ (conventional unit cell). 
(b) Bulk Brillouin zone (BZ) and the projected (001) surface BZ, with high-symmetry points indicated. 
(c) Electronic band structure calculated using the modified Becke-Johnson (mBJ) exchange potential.}
\end{figure}

\textit{\textcolor{black}{Results. ---}}
As shown in Fig.~\ref{fig-1}(a), $\beta$-Bi$_4$I$_4$ is a typical quasi-one-dimensional van der Waals material that crystallizes in the monoclinic space group $C2/m$ (No. 12). 
Its building blocks are infinite Bi$_4$I$_4$ molecular chains that extend along the crystallographic $b$-axis. 
Internally, each chain consists of a central bismuth bilayer with a zigzag atomic arrangement, which is terminated by iodine atoms through strong covalent bonds. 
These one-dimensional chains are packed together along the $a$ and $c$ axes and are held together by weak van der Waals interactions. 
Due to this highly anisotropic stacking, the material typically forms needle-like single crystals and possesses two distinct natural cleavage planes, the (001) top surface and the (100) side surface.
$\beta$-Bi$_4$I$_4$ has been experimentally synthesized, and its energy bands have been verified by angle-resolved photoemission spectroscopy (ARPES) measurements ~\cite{autes2016novel,noguchi2019weak,zhao2024topological}. In the conventional cell of $\beta$-Bi$_4$I$_4$, as depicted in Fig.~\ref{fig-1}(a), the experimental lattice constants read $ a = 14.386 \ \text{\AA}, b = 4.430 \ \text{\AA}, c = 10.493 \ \text{\AA} \text{, and} \ \beta = 107.87^{\circ} $~\cite{von1978kenntnis}.

As illustrated in Fig.~\ref{fig-1}(c), the SOC-included band structure exhibits a narrow gap along the high-symmetry lines of the Brillouin zone [see Fig.~\ref{fig-1}(b)].
The classification of the band topology in $\beta$-Bi$_4$I$_4$ is sensitive to calculation methods and external conditions such as strain and temperatures~\cite{autes2016novel,liu2016weak,noguchi2019weak,zhao2024topological,huang2021room,hu2025phonons}.
Among eight time-reversal invariant points, the SOC-driven band inversions may occur at both the $M$ and $L$ points, which places the material in the proximity of the phase boundary between a weak topological insulator (WTI) [$\mathbb{Z}_2$ indices $(0;001)$] and a strong topological insulator (STI) [$\mathbb{Z}_2$ indices $(1;110)$] or the phase boundary between an STI and a normal insulator (NI) [$\mathbb{Z}_2$ indices $(0;000)$]. 
By employing the mBJ exchange potential, we identify $\beta$-Bi$_4$I$_4$ as a NI. As shown in Fig.~\ref{fig-1}(c), the equilibrium band structure exhibits a narrow direct band gap of $\simeq11$~meV at the high-symmetry point $M$. 
While our results yield a topology distinct from the experimental WTI assignment~\cite{noguchi2019weak}, theoretical predictions suggest that the material undergoes a transition to a normal insulator under minute tensile strain~\cite{liu2016weak}. 

\begin{figure}[t!]
\includegraphics[width=1\columnwidth]{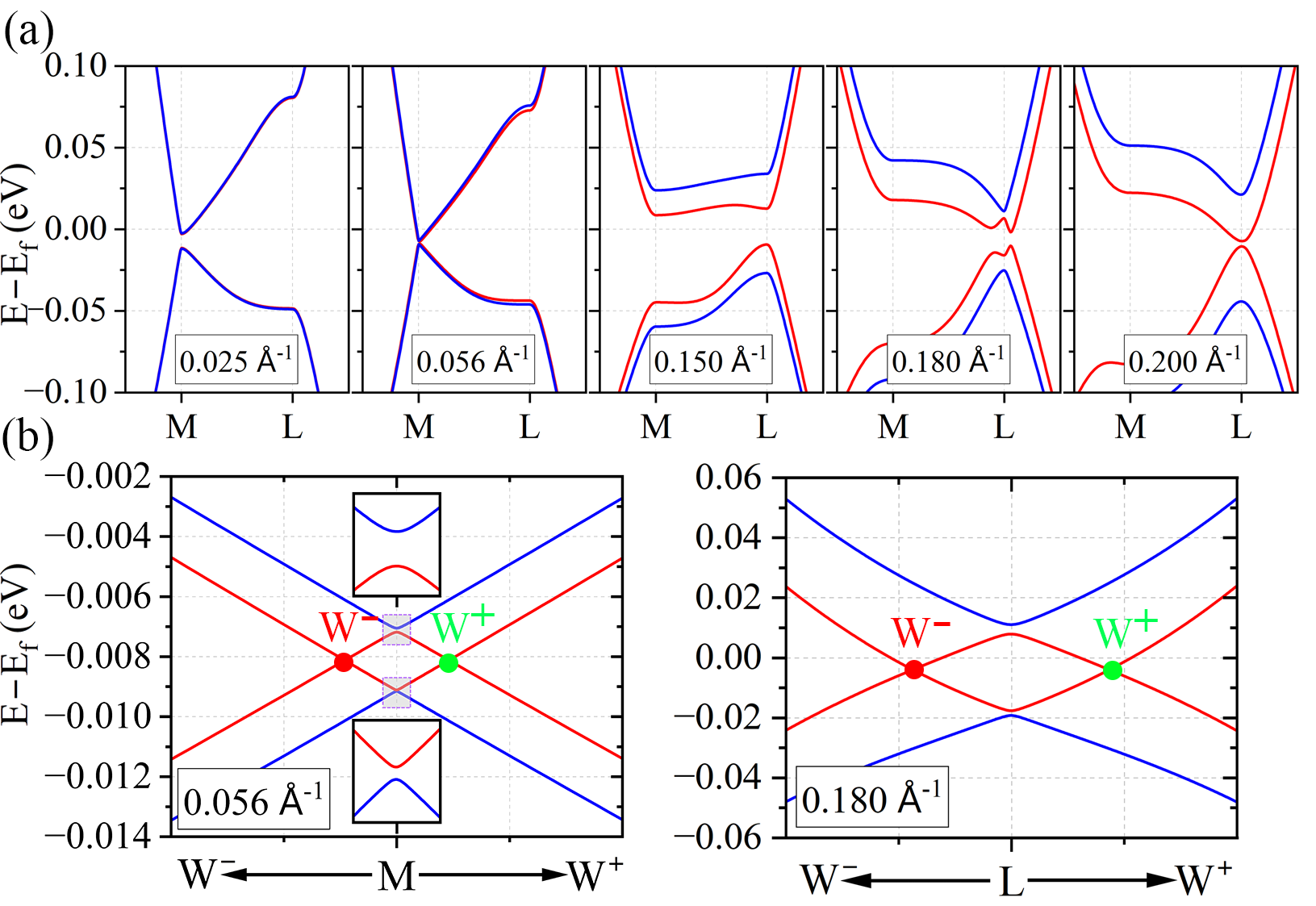}
\caption{\label{fig-2}(a) Evolution of the band structure of $\beta$-Bi$_4$I$_4$ under circularly polarized light with increasing light intensity $\mathrm{e}\mathrm{A}_{0}/\hbar$. 
(b) Band dispersions at $\mathrm{e}\mathrm{A}_{0}/\hbar = 0.056$ and $0.180$~\AA$^{-1}$, respectively. 
Regarding the color coding: under optical driving, each original band splits into two branches. The red lines denote the inner branches that approach and intersect to form Weyl nodes, while the blue lines denote the outer branches that move away. The inset in the left panel of (b) confirms that there is no band crossing at the $M$ point.
In these gapless phases, the system hosts a single pair of Floquet Weyl nodes $W^{\pm}$, located near the $M$ point and the $L$ point. Here, the superscript $\pm$ denotes the chirality of the Weyl node is $\pm1$.
The Weyl nodes with positive and negative chirality are marked in red and green, respectively. }
\end{figure}

Then, we consider the system under the influence of a spatially homogeneous periodic driving field propagating along the $z$-axis. The corresponding vector potential, which is polarized within the $x$-$y$ plane, is given by $\mathbf{A}(t) = \mathrm{A}_{0}[\cos(\omega t), \sin(\omega t+\varphi), 0]$ with $\varphi=0$ corresponding to CPL.
In our calculations, we set the photon energy to $\hbar\omega=10$~eV to satisfy the requirements of high-frequency approximation.
As shown in Fig.~\ref{fig-2}(a), under circularly polarized light irradiation, the spin degeneracy of the energy bands is lifted due to the breaking of time-reversal symmetry.
To systematically investigate the influence of light intensity on the topological phase transitions in $\beta$-Bi$_4$I$_4$, the light intensity, parameterized by $\mathrm{e}\mathrm{A}_{0}/\hbar$ in units of \AA$^{-1}$, was varied gradually from $0.00$~\AA$^{-1}$ to $0.22$~\AA$^{-1}$, while the photon energy was fixed.
Under left-handed CPL irradiation, we systematically investigated the evolution of the band structure of $\beta$-Bi$_4$I$_4$. 
The evolution of the band structure is shown in Fig.~\ref{fig-2}(a) and the Supplementary Material (SM)~\cite{Hwang}. 
The application of a light field was found to induce a significant modulation in the electronic bands of $\beta$-Bi$_4$I$_4$, reflecting the effective light–matter interaction in this system.
As shown in Fig. S1 and Fig. S3 in the SM~\cite{Hwang}, within the light intensity range of $0.00$~\AA$^{-1}$ to $0.22$~\AA$^{-1}$, the system undergoes two successive band inversion processes (gap closing and reopening). 
Specifically, within the range of $0.052-0.061$~\AA$^{-1}$ and $0.165-0.202$~\AA$^{-1}$, the CPL-irradiated $\beta$-Bi$_4$I$_4$ is driven into a gapless phase. Outside these two parameter ranges, the CPL-irradiated $\beta$-Bi$_4$I$_4$ exhibits a global energy gap.

\begin{figure}[t]
\includegraphics[width=1\columnwidth]{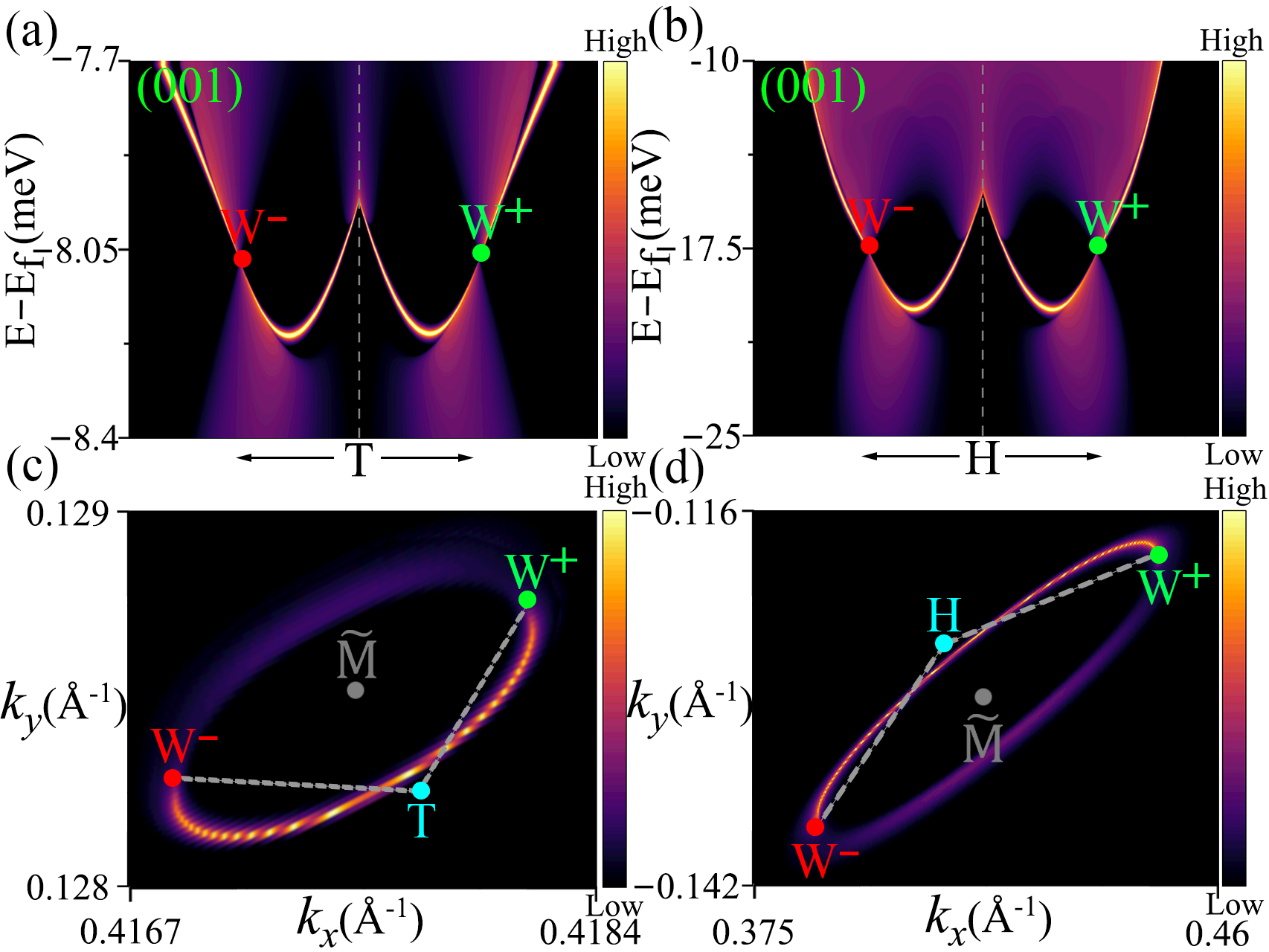}
\caption{\label{fig-3}(a), (b) Surface spectrum on the (001) surface of $\beta$-Bi$_4$I$_4$ driven by circularly polarized light. 
The field amplitudes are $eA_{0}/\hbar = 0.054$ and $0.186$~\AA$^{-1}$, respectively.
(c), (d) Corresponding isoenergy contours on the (001) surface calculated at the energy midpoint between the Weyl nodes.
Red and green dots mark the projected Weyl nodes with positive and negative chirality, respectively.
The light blue dots labeled $T$ and $H$ in (c) and (d) indicate the specific momenta where the surface spectrum in (a) and (b) were sampled. The lines dotted in gray denote the routes shown in (a), (b).}
\end{figure}

Based on first-principles calculations, we perform a detailed investigation of the gapless phases in CPL-irradiated $\beta$-Bi$_4$I$_4$ by selecting two representative light intensities, $\mathrm{e}\mathrm{A}_{0}/\hbar=0.056$~\AA$^{-1}$ and $0.180$~\AA$^{-1}$. 
Our results demonstrate that in both cases, the system manifests as a Floquet Weyl semimetal hosting a single pair of Weyl nodes. 
Specifically, as shown in Fig.~\ref{fig-2}(b) and Fig. S3 in the SM~\cite{Hwang}, for $\mathrm{e}\mathrm{A}_{0}/\hbar=0.056$~\AA$^{-1}$, the Weyl nodes with opposite chirality emerge near the high-symmetry $M$ point, whereas for $\mathrm{e}\mathrm{A}_{0}/\hbar=0.180$~\AA$^{-1}$, they are located in the vicinity of the $L$ point.
To further substantiate the non-trivial topology of the Floquet Weyl semimetal phase in CPL-irradiated $\beta$-Bi$_4$I$_4$, we calculate the surface states for light intensities $\mathrm{e}\mathrm{A}_{0}/\hbar = 0.054$ and $0.180$~\AA$^{-1}$ using the iterative Green's function method based on the 
effective static Hamiltonian obtained by the Floquet theorem. 
The resulting local density of states (LDOS) reveals topological surface states that terminate at the projected Weyl nodes $W^{\pm}$, as shown in Figs.~\ref{fig-3}(a) and (b). 
As illustrated in Figs.~\ref{fig-3}(c) and (d), the corresponding Fermi arcs connect pairs of Weyl nodes with opposite chirality ($\pm 1$), serving as a definitive fingerprint of the Floquet Weyl semimetal state. 

Next, by setting the phase parameter $\varphi = \pi/2$, the driving field becomes LPL with polarization along the $[1,-1,0]$ direction. 
It is important to note that, unlike CPL, LPL preserves both time-reversal and spatial inversion symmetries. 
Consequently, the spin degeneracy of $\beta$-Bi$_4$I$_4$ remains intact, and the system cannot be driven into a Floquet Weyl semimetal phase. 
The evolution of the energy gap under LPL irradiation is shown in detail in Fig. S1 and Fig. S2 of the SM~\cite{Hwang}. 
Our calculations reveal that, within the light intensity range of $\mathrm{e}\mathrm{A}_{0}/\hbar=0.00\mapsto0.22$~\AA$^{-1}$, the LPL-irradiated $\beta$-Bi$_4$I$_4$ undergoes a sequence of topological phase transitions, evolving from a normal insulator [$\mathbb{Z}_2$ indices $(0;000)$] to a strong topological insulator [$\mathbb{Z}_2$ indices $(1;110)$], and subsequently to a weak topological insulator [$\mathbb{Z}_2$ indices $(0;001)$]. 
These two topological phase transitions occur at the critical light intensities of $\mathrm{e}\mathrm{A}_{0}/\hbar\simeq0.056$~\AA$^{-1}$ and $0.168$~\AA$^{-1}$, respectively.

\begin{figure}[t]
\includegraphics[width=1\columnwidth]{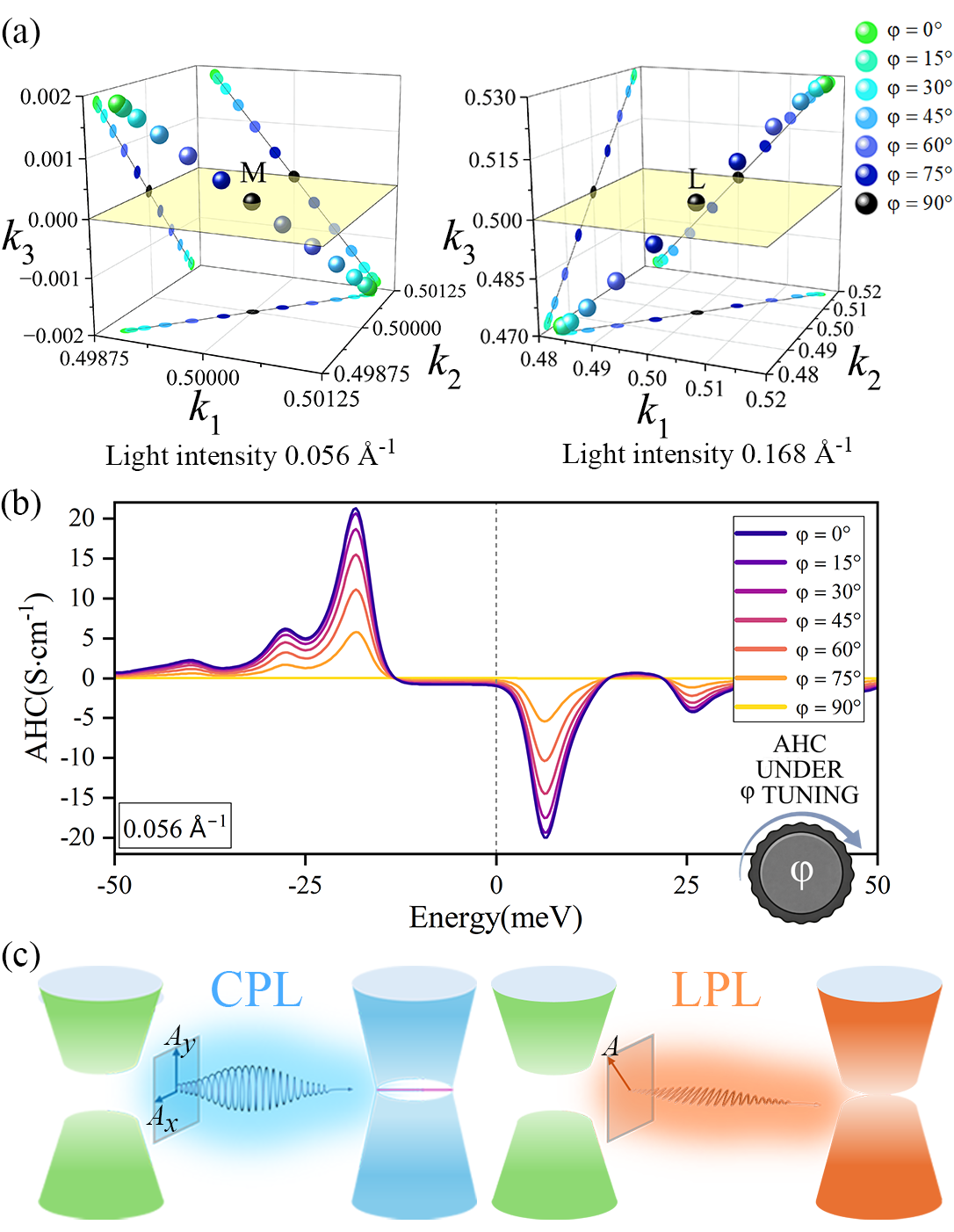}
\caption{\label{fig-4}(a) Trajectories of Floquet Weyl nodes in reciprocal space ($k_{1}$-$k_{3}$) driven by light with a tunable phase difference $\varphi$. 
The left and right panels correspond to field amplitudes $eA_{0}/\hbar = 0.056$ and $0.168$~\AA$^{-1}$, respectively. 
$\varphi$ defines the phase delay between the orthogonal linear components. 
Solid black dots mark the merging points of Weyl nodes with opposite chirality at high-symmetry $M$ or $L$ points when $\varphi = 90^{\circ}$.
(b) Calculated anomalous Hall conductivity $\sigma_{xy}$ as a function of chemical potential for various $\varphi$ in the light intensity of $eA_{0}/\hbar = 0.056$~\AA$^{-1}$. 
Note that $\sigma_{xy}$ decreases and eventually vanishes as $\varphi$ increases.
(c) Schematic phase diagram illustrating the topological transition driven by the polarization state of the light.}
\end{figure}

Motivated by these observations, we fix the light intensity at a value where the CPL-irradiated $\beta$-Bi$_4$I$_4$ resides in the Floquet Weyl semimetal phase, i.e., within the parameter range of $0.052-0.061$~\AA$^{-1}$ or $0.165-0.202$~\AA$^{-1}$. 
By continuously tuning the phase difference between the two orthogonal linear polarization components $\varphi$ from $0$ to $\pi/2$, we are able to trace the creation or annihilation processes of the Weyl nodes.
As depicted in Fig.~\ref{fig-4}(a), for a fixed light intensity of $\mathrm{e}\mathrm{A}_{0}/\hbar=0.056$ ($0.168$)~\AA$^{-1}$, increasing $\varphi$ causes the Weyl nodes within each pair to approach one another. 
Upon reaching $\varphi = \pi/2$, nodes with opposite chirality merge at the high-symmetry $M$ ($L$) point to form a Dirac fermion, thereby extinguishing the nontrivial Floquet Weyl semimetal phase.

Intriguingly, beyond tracking the creation and annihilation of Floquet Weyl nodes, the phase difference $\varphi$ provides an effective means to modulate the anomalous Hall conductivity, and even effectively acts as a switch for the anomalous Hall effect in light-irradiated $\beta$-Bi$_4$I$_4$.
The intrinsic anomalous Hall conductivity $\sigma_{xy}$ is calculated by integrating the Berry curvature over the Brillouin zone~\cite{yao2004first},
\begin{equation}
    \sigma_{xy} = -\frac{e^2}{\hbar} \int_{\mathrm{BZ}} \frac{d^3k}{(2\pi)^3} \sum_{n} f(E_{n\mathbf{k}}) \Omega_{n,z}(\mathbf{k}),
    \label{eq:AHC_Berry}
\end{equation}
where $f(E_{n\mathbf{k}})$ is the Fermi-Dirac distribution function for the $n$-th band with energy $E_{n\mathbf{k}}$, and $\Omega_{n,z}(\mathbf{k})$ is the $z$-component of the Berry curvature. In a three-dimensional Weyl semimetal phase, provided that time-reversal symmetry is broken, the existence of a pair of Weyl nodes with opposite chirality separated by a distance $\Delta k_z$ along the $k_z$-axis generates an intrinsic anomalous Hall conductivity proportional to the separation as~\cite{burkov2014anomalous},
\begin{equation}
    \sigma_{xy} = \frac{e^2}{4\pi^2} \Delta
    k_z.
    \label{eq:AHC_Weyl}
\end{equation}
Taking light-irradiated $\beta$-Bi$_4$I$_4$ as an example, for the case of $\mathrm{e}\mathrm{A}_{0}/\hbar=0.056$ ~\AA$^{-1}$, $\varphi-\pi/2\neq0$ indicates the breaking of time-reversal symmetry and nonzero anomalous Hall conductivity. 
When $\varphi-\pi/2=0$, the preservation of time-reversal symmetry results in the vanishing of the anomalous Hall conductivity due to $\Omega_{n,z}(\mathbf{k})=-\Omega_{n,z}(-\mathbf{k})$. Thus, as shown in Fig.~\ref{fig-4}(b), the increase of $\varphi$ reflects the decrease of anomalous Hall conductivity. Furthermore, the anomalous Hall conductivity eventually becomes zero when $\varphi=\pi/2$.
Thus, we propose that the anomalous Hall conductivity would be an effective means of tracking the Floquet Weyl nodes in $\beta$-Bi$_4$I$_4$.

\textit{\textcolor{black}{Summary and Discussions. ---}}
In summary, we have demonstrated that breaking time-reversal symmetry with circularly polarized light drives quasi-one-dimensional $\beta$-Bi$_4$I$_4$ into a Floquet WSM phase [as schematically shown in Fig.~\ref{fig-4}(c)]. 
Within this phase, the phase difference $\varphi$ between the two orthogonal linear polarization components acts as a precise control parameter. 
Specifically, continuous tuning of $\varphi$ governs the trajectory and eventual annihilation of Weyl nodes near high-symmetry points ($L$ and $M$). 
Crucially, this topological evolution is directly mapped onto a continuous change in the anomalous Hall conductivity, providing a unique, all-electrical pathway to track the light-induced phase transition. 
Compared to traditional Floquet engineering that relies on modulating light intensity or frequency, our $\varphi$-driven mechanism offers a significantly more accessible and precise experimental route for topological switching. 
To firmly establish the viability of this proposed route, we have detailed the effects of varying polarization handedness and incidence angles, alongside a comprehensive multi-perspective analysis of the experimental feasibility, in the SM~\cite{Hwang}. 
Furthermore, owing to its weak inter-chain coupling and natural proximity to multiple topological phase boundaries, $\beta$-Bi$_4$I$_4$ emerges as an ideal and realistic material platform for such optically controlled topological electronics. 
Inspired by the recent experimental detection of the light-induced anomalous Hall effect in graphene~\cite{mciver2020light}, our findings are expected to stimulate further investigations in ultrafast spectroscopy and material design, ultimately laying the groundwork for high-speed, light-controlled quantum devices.

\textit{\textcolor{black}{Acknowledgments. ---}}
This work was supported by the National Key Research and Development Program of the Ministry of Science and Technology of China (Grant No. 2025YFA1411303), the National Natural Science Foundation of China (NSFC, Grants No. 12547101 and No. 92365101), the Natural Science Foundation of Chongqing (Grants No. 2023NSCQ-JQX0024 and No. CSTB2022NSCQMSX0568), the Beijing National Laboratory for Condensed Matter Physics (Grant No. 2024BNLCMPKF015), and the Open Research Fund of State Key Laboratory of Quantum Functional Materials (NO. QFM2025KF002).

\bibliography{reference}

\end{document}